\documentclass[prd,twocolumn,floatfix,preprintnumbers]{revtex4}

\usepackage{graphicx}
\input{epsf}
\usepackage{epsf}
\usepackage{graphicx,epsfig}
\usepackage{bm}
\usepackage{latexsym,amssymb,amsmath,float}

\newcommand {\ga} {\ {\raise-.5ex\hbox{$\buildrel>\over\sim$}}\ }
\newcommand {\la} {\ {\raise-.5ex\hbox{$\buildrel<\over\sim$}}\ } 
\newcommand{\eqn}[1] {Eq.~(\ref{#1})}

\def\be{\begin{equation}}
\def\ee{\end{equation}}
\def\ba{\begin{eqnarray}}
\def\ea{\end{eqnarray}}

\renewcommand{\(}{\left(} 
\renewcommand{\)}{\right)} 
\renewcommand{\[}{\left[} 
\renewcommand{\]}{\right]}

\begin{document}

\title{Decaying dark matter mimicking time-varying dark energy}
\author{Sourish Dutta and Robert J. Scherrer}
\affiliation{Department of Physics and Astronomy, Vanderbilt University,
Nashville, TN  ~~37235}

\begin{abstract}
A $\Lambda$CDM model with dark matter that decays into inert relativistic
energy on a timescale longer than the Hubble time will produce an
expansion history that can be misinterpreted as stable dark matter with
time-varying dark energy.  We calculate the corresponding spurious equation
of state parameter, $\widetilde w_\phi$, as a function of
redshift, and show that the evolution
of $\widetilde w_\phi$ depends strongly on the assumed value of the
dark matter density, erroneously taken to scale as $a^{-3}$.  Depending on the latter,
one can obtain models that mimic quintessence ($\widetilde w_\phi > -1$),
phantom models ($\widetilde w_\phi < -1$) or models in which
the equation of state parameter crosses the phantom
divide, evolving from $\widetilde w_\phi > -1$ at high redshift to
$\widetilde w_\phi < -1$ at low redshift.  All of these models generically
converge toward $\widetilde w_\phi \approx -1$ at the present.  The degeneracy between
the $\Lambda$CDM model with decaying dark matter and the corresponding spurious
quintessence model is broken by the growth of density perturbations.
\end{abstract}

\maketitle

\section{Introduction}

Cosmological data from a wide range of sources including type Ia supernovae \cite{hicken}, the cosmic microwave background \cite{Komatsu:2010fb}, baryon acoustic oscillations \cite{percival}, cluster gas fractions \cite{Samushia2007,Ettori} and gamma ray bursts \cite{Wang,Samushia2009} seem to indicate that at 
least 70\% of the energy density in the
universe is in the form of an exotic, negative-pressure component,
called dark energy.  (See Ref. \cite{Copeland} for a comprehensive
review).  A parameter of considerable importance is the equation of state (EoS) of the dark energy component, defined as the ratio of its pressure to its density:
\be
w=p_{\rm DE}/\rho_{\rm DE}.
\ee
If the dark energy is due to a cosmological constant, then $w$ is constant
and exactly equal to $-1$.  A value of $w=-1$ is consistent with current
observations \cite{Wood-Vasey,Davis}.
On the other hand, a variety of models
have been proposed in which $w$ is time varying.  Perhaps
the simplest model for time-varying $w$ is to take
the dark energy to be due to
a scalar field, dubbed ``quintessence"
\cite{ratra,wett88,turner,caldwelletal,liddle,Stein1,DS,Dutta1,Dutta2}.
As the data continue
to improve, it is important to be able to distinguish models (such as
quintessence) with a true
time-varying $w$ from $\Lambda$CDM models which can effectively
mimic such a time variation.

To motivate our investigation, consider first the simple $\Lambda$CDM model,
with present-day dark matter and cosmological constant densities
of $\rho_{DM0}$ and $\rho_{\Lambda0}$,
respectively.  (Zero subscripts will be used throughout to denote
present-day values of cosmological quantities).
If one does not know {\it a priori} the value of $\rho_{DM0}$
or the fact that the dark energy is
a pure cosmological constant, then part of the dark matter
density can be absorbed into the dark energy, producing (erroneously)
a time-varying dark energy.  In particular,
the $\Lambda$CDM model with $\rho_{DM0}$, $\rho_{\Lambda0}$ is degenerate
with the quintessence model having a present-day dark matter density
$\widetilde \rho_{DM0} \ne \rho_{DM0}$, and a
time-varying dark energy component
with density $\widetilde \rho_\phi = \rho_{\Lambda} + \rho_{DM} -
\widetilde \rho_{DM}$ and equation of state
\begin{equation}
\label{wdm}
\widetilde w_\phi = -\frac{\rho_{\Lambda0}}{\rho_{\Lambda0} +
(\rho_{DM0} - \widetilde \rho_{DM0})(1+z)^3.}
\end{equation}
(Since our paper deals with spurious measurements of a time-varying
equation of state, we will use tildes throughout to deal with
spurious/unphysical/mismeasured
quantities; non-tilded variables will refer to true physical quantities). 
This degeneracy has been exhaustively explored in Refs.
\cite{Hu,Wasserman,Rubano,Kunz}, who pointed out that it cannot be resolved without
independent knowledge of the dark matter density.

On the other hand, if observational data led to an equation
of state for the dark energy that mimicked the evolution of $w_\phi$ given in Eq. (\ref{wdm}),
Occam's razor would suggest that the correct model for the universe was
actually $\Lambda$CDM, with the appropriately different value for
the dark matter density.
It is of interest to determine if
there are any less obvious spurious time-varying
equations of state for the dark energy that arise from
simple variations on $\Lambda$CDM.

One such model is $\Lambda$CDM with decaying dark matter.
If the dark matter decays into inert radiation, and one interprets
the measured behavior of $H(z)$ under the
incorrect assumption that the dark
matter is stable, then the best fit to the observations will
be a quintessence model with a time-varying equation of state.  This
occurs because the nonstandard time-variation in both the
dark matter and inert radiation densities gets absorbed into
time variation of the dark energy.
This was first noted by Ziaeepour \cite{Z}, who argued that it leads, in
general, to a mismeasured value of the equation of state parameter
satisfying $\widetilde w_\phi < -1$.  (Note that
similar $\widetilde w_\phi < -1$ behavior occurs when energy can be exchanged between
dark matter and dark energy \cite{Das}, but this is a different class of models
than those discussed here).

In this paper we reexamine the behavior of such models, and provide
an improved calculation of $\widetilde w_\phi(z)$.  We show that the
behavior of $\widetilde w_\phi(z)$ is very sensitive to the assumed
value of $\widetilde \rho_{DM}$.  Different choices for $\widetilde \rho_{DM}$
can lead to models that mimic quintessence, phantom-like models, or models
that cross the phantom divide.
In the next section, we present the calculation
of $\widetilde w_\phi(z)$, and we discuss our results in Sec. III.

\section{The Effective Equation of State}

We consider a $\Lambda$CDM Universe which consists of
baryonic matter with density $\rho_B$,
dark matter with density $\rho_{DM}$, and cosmological constant with density
$\rho_\Lambda$.  The dark matter is assumed to decay at a rate $\Gamma$
into invisible relativistic energy with density $\rho_R$.  By ``invisible", we
mean that the relativistic decay products are assumed to interact very weakly
with ordinary matter.  If the decay products did interact, e.g.,
electromagnetically, they would
be easily detectable at the lifetimes of interest here, and the model
would already be ruled out.

The equations for the evolution of matter and the decay-produced radiation are the following:
\ba
\label{rhoM_eom}
\dot{\rho}_{DM}&=&-3H\rho_{DM}-\Gamma\rho_{DM},\\
\label{rhoR_eom}
\dot{\rho}_R&=&-4H\rho_{R}+\Gamma\rho_{DM},\\
\label{rhoB_eom}
\dot{\rho}_B&=&-3H\rho_B,
\ea
where $H$ is the Hubble parameter given by the Friedman equation
\be
\label{Friedman_DDM}3H^2=8\pi G\(\rho_{DM}+\rho_B+\rho_R+\rho_\Lambda\),
\ee
and dots denote time derivatives.
We ignore ``ordinary'' radiation,
since it has a negligible effect on
the expansion rate in the redshift regime relevant to this problem.

The existence of such decays with lifetimes comparable to the Hubble time
can be constrained by their effects on the CMB, large-scale structure,
and SN Ia observations \cite{Kaplinghat,Ichiki,Gong,Amigo}.  We will confine our attention
to lifetimes satisfying the constraint given
in Ref. \cite{Amigo}: $\Gamma^{-1} > 100$ Gyr, or $\Gamma t_0 < 0.15$, where $t_0$
is the present-day age of the universe.

The $\rho_R$ term in the expression for $H$, as well as the nonstandard evolution
of $\rho_{DM}$, will clearly
lead to an effective equation of state different from the $\Lambda$CDM cosmology.
Observers unaware of the decaying nature of the dark matter might try
to explain the time-varying equation of state with the help of a
quintessence field $\phi$ in addition to ordinary non-decaying dark matter.
In their model, (denoting the density of presumed
non-decaying dark matter by $\widetilde\rho_{DM}$) the same
expansion rate $H$ will be given by:
\be
\label{Friedman_phi}3H^2=8\pi G\(\widetilde\rho_{DM}+\rho_B +
\widetilde\rho_\phi\).
\ee

Equating \eqn{Friedman_DDM} and \eqn{Friedman_phi}, one can readily deduce the energy density of this fictitious scalar field in terms of the densities of matter and decay-produced radiation as:
\be
\label{rho_phi}
\widetilde\rho_{\phi}=\rho_{\Lambda}+\rho_R+\rho_{DM}-\widetilde\rho_{DM}.
\ee
From this, it is straightforward to determine the effective equation of state of
this fictitious dark energy component from the relation $-3\(1+\widetilde
w_\phi\)=d\ln\widetilde
\rho_\phi/d\ln a$ (and also using
equations \eqref{rhoM_eom} and \eqref{rhoR_eom}):
\be
\label{w}
\widetilde w_{\phi}=\frac{\rho_{R}/3 - \rho_\Lambda}
{\rho_{\Lambda}+\rho_R+\rho_{DM}-\widetilde\rho_{DM}}.
\ee
Note that $\widetilde w_\phi$ does not depend explicitly on $\rho_B$.  However,
there is an implicit dependence, since we assume a flat geometry with
$\Omega=1$, so that $\Omega_{DM} + \Omega_\Lambda + \Omega_R = 1 - \Omega_B$.
(In this paper, we will use $\Omega_{DM}$, $\Omega_\Lambda$,... to refer only to present-day values, so
we drop the zero subscript in these cases).

Consider first the qualitative behavior of $\widetilde w_\phi$.
The solutions to Eqs. (\ref{rhoM_eom}) and (\ref{rhoR_eom})
can be written
in terms of the scale factor $a(t)$ as \cite{ST} 
\ba
\label{rhoM_sol}
\rho_{DM}&=&\rho_{DM0}\(\frac{a}{a_0}\)^{-3}\exp\[-\Gamma (t-t_0)\]\\
                 \rho_R&=&\rho_{DM0}\(\frac{a}{a_0}\)^{-4}\times\nonumber\\
\label{rhoR_sol} &&\int_{0}^{t}\(\frac{a(t')}{a_0}\)\exp\(-\Gamma t'\)\Gamma dt',
\ea
while the fictitious non-decaying dark matter density evolves, of course, as  
\be
\label{rhom_sol}
\widetilde \rho_{DM}=\widetilde\rho_{DM0}\(\frac{a}{a_0}\)^{-3}.
\ee

Clearly, the behavior of $\widetilde w_\phi$ depends on the assumed
value of $\widetilde\rho_{DM0}$.
Using SN Ia data alone (or more generally,
the behavior of $H(z)$ alone) there is no best-fit
value for the dark matter density,
once the equation of state of the dark energy is allowed to be a free
function of $z$, just as in the
case of non-decaying dark matter discussed
earlier \cite{Hu,Wasserman,Rubano,Kunz}, and the best
one can do is to derive the
behavior of $\widetilde w_\phi$ as a function of $\widetilde \rho_{DM0}$.
Of course, there are
other cosmological measurements that constrain the dark matter density.
For instance, large-scale structure provides
a constraint on the present-day dark matter density, while the CMB
constrains the dark matter density at high redshift, but taken together
these limits are consistent with a change in the comoving dark matter density
as large as 15\%, as noted previously \cite{Amigo}.
However, it is reasonable to assume that an observer erroneously
postulating stable dark matter would derive
a value for $\widetilde \rho_{DM0}$ that lies somewhere between the
values obtained by taking $\widetilde \rho_{DM} = \rho_{DM}$ at high
redshift ($\Gamma t \ll 1$) or by taking $\widetilde \rho_{DM} = \rho_{DM}$
at the present day ($t = t_0$).
Equating $\rho_{DM}$ from Eq. (\ref{rhoM_sol}) and $\widetilde \rho_{DM}$
from Eq. (\ref{rhom_sol}) at $t \rightarrow 0$ and $t = t_0$ then gives
the plausible bounds on $\widetilde \rho_{DM0}$:
\begin{equation}
\rho_{DM0} \le \widetilde \rho_{DM0} \le \rho_{DM0} \exp(\Gamma t_0),
\end{equation}
where the upper bound corresponds to equality between
$\widetilde \rho_{DM}$ and $\rho_{DM}$
at high redshift, while the lower
bound assumes equality today.

To parametrize this uncertainty, we will take
\begin{equation}
\label{Deltadef}
\widetilde \rho_{DM0} = \rho_{DM0} \exp(\Delta \Gamma t_0),
\end{equation}
where $0 \le \Delta \le 1$.  Here $\Delta = 0$ corresponds to setting the
(spurious) stable dark matter density equal to the decaying dark matter density at the
present, while $\Delta = 1$ corresponds to equality $t \rightarrow 0$.

To gain some qualitative insight into the behavior of $\widetilde w_\phi$, it
is necessary to derive an approximation to the decay-produced radiation density
in Eq. (\ref{rhoR_sol}). 
Note that we always have $\exp(-\Gamma t) \approx 1- \Gamma t$
for our observational bound of $\Gamma t_0 < 0.15$.  This approximation
can be used to integrate Eq. (\ref{rhoR_sol})
in the matter dominated era ($z \ga 1$)
to yield \cite{ST}
\begin{equation}
\label{rhoR_approx}
\rho_R = \frac{3}{5} \Gamma t \rho_{DM0} (a/a_0)^{-3}.
\end{equation}
This result becomes progressively less accurate as the cosmological constant begins to dominate at late times,
but it will be sufficient to provide qualitative insight into the
evolution of $\widetilde w_\phi$.

Substituting the expressions for $\rho_{DM}$, $\widetilde \rho_{DM}$ and
$\rho_R$ from Eqs. (\ref{rhoM_sol}), (\ref{rhom_sol}) and
(\ref{rhoR_approx}), respectively, into Eq. (\ref{w}), and expanding
out to lowest order in $\Gamma t$ and $\Gamma t_0$ (both of which
are $\ll 1$), we obtain
\begin{equation}
\label{w_ana}
\widetilde w_\phi = \frac{-\rho_\Lambda + \frac{1}{5} \Gamma t\rho_{DM0} (1+z)^3}
{\rho_\Lambda + [(1-\Delta) \Gamma t_0 - \frac{2}{5} \Gamma t]\rho_{DM0}
(1+z)^3}.
\end{equation}

In order to express $\widetilde w_\phi$ as a function of redshift
alone, we make
the approximation that the expansion law
never diverges very far from $\Lambda$CDM (see, e.g.,
Ref. \cite{DS}).
In this case, the cosmic time $t$ is related to the redshift as
\be
\label{tz}
t(z)=t_\Lambda \sinh^{-1}\[\frac{[\Omega_\Lambda/(1-\Omega_\Lambda)]^{1/2}}
{\(1+z\)^{3/2}}\],
\ee
where $t_\Lambda$ is a constant with units of time that is
related to the energy density of the cosmological constant by
\be
t_\Lambda=\sqrt{1/(6\pi G\rho_\Lambda)}.
\ee
Eq. (\ref{tz}) will be a good approximation as long as $\widetilde w_\phi$
is close to $-1$ whenever the dark energy dominates the expansion.
In the models we investigate here, $w_\phi$ can significantly
diverge from $-1$
only at early times, when the expansion is dark-matter dominated, so
we expect Eq. (\ref{tz}) to be sufficiently accurate for our purposes.
Then Eq. (\ref{w_ana}) can be written as
\begin{align}
\label{wfinal}
&\widetilde w_\phi = \nonumber\\
&\frac{-\Omega_\Lambda + (1/5)\Gamma t_0 f(z) (1-\Omega_\Lambda)(1+z)^3}
{\Omega_\Lambda + [(1-\Delta)  - (2/5) f(z)]\Gamma t_0
(1-\Omega_\Lambda)(1+z)^3}.
\end{align}
where $f(z)$ is the fractional age of the universe at redshift $z$, given by
\be
\label{fz}
f(z)\equiv \frac{t(z)}{t_0} =
\frac{\sinh^{-1}\[\[\Omega_\Lambda/(1-\Omega_\Lambda)\]^{1/2}(1+z)^{-3/2}\]}
{\tanh^{-1}\(\sqrt{\Omega_\Lambda}\)}.
\ee

In Figs. 1-2, we show the exact evolution of $\widetilde w_\phi$, derived
by numerical integration of Eqs. \eqref{rhoM_eom} and \eqref{rhoR_eom}, along
with our approximation given by Eqs. (\ref{wfinal}) and (\ref{fz}), for
different values of $\Gamma t_0$, $\Delta$, and $\Omega_\Lambda$.  We
see that our analytic approximation is quite accurate for all of the cases
examined.
\begin{center}
\begin{figure*}[!]
\begin{tabular}{c@{\qquad}c}
\epsfig{file=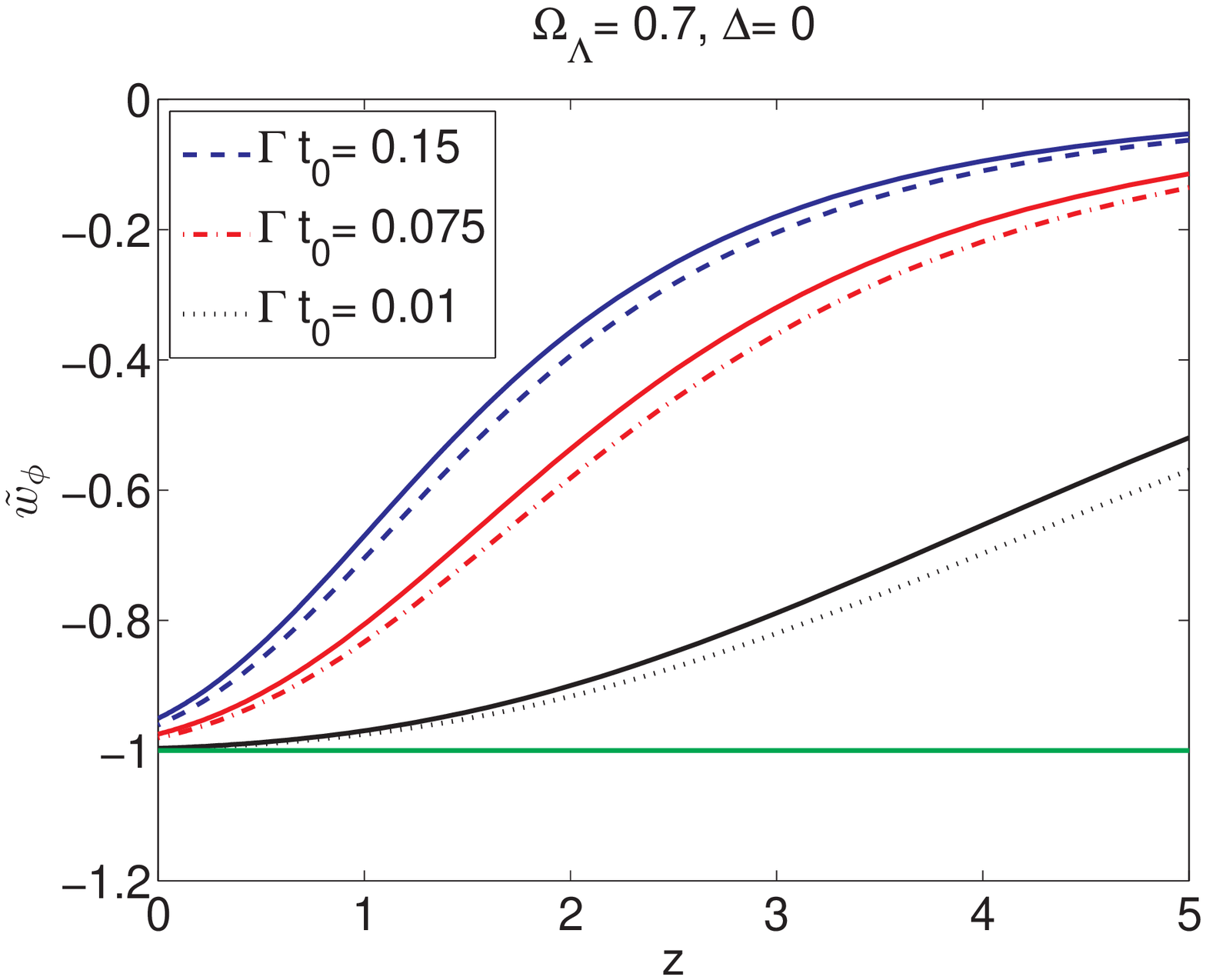,width=7 cm}&\epsfig{file=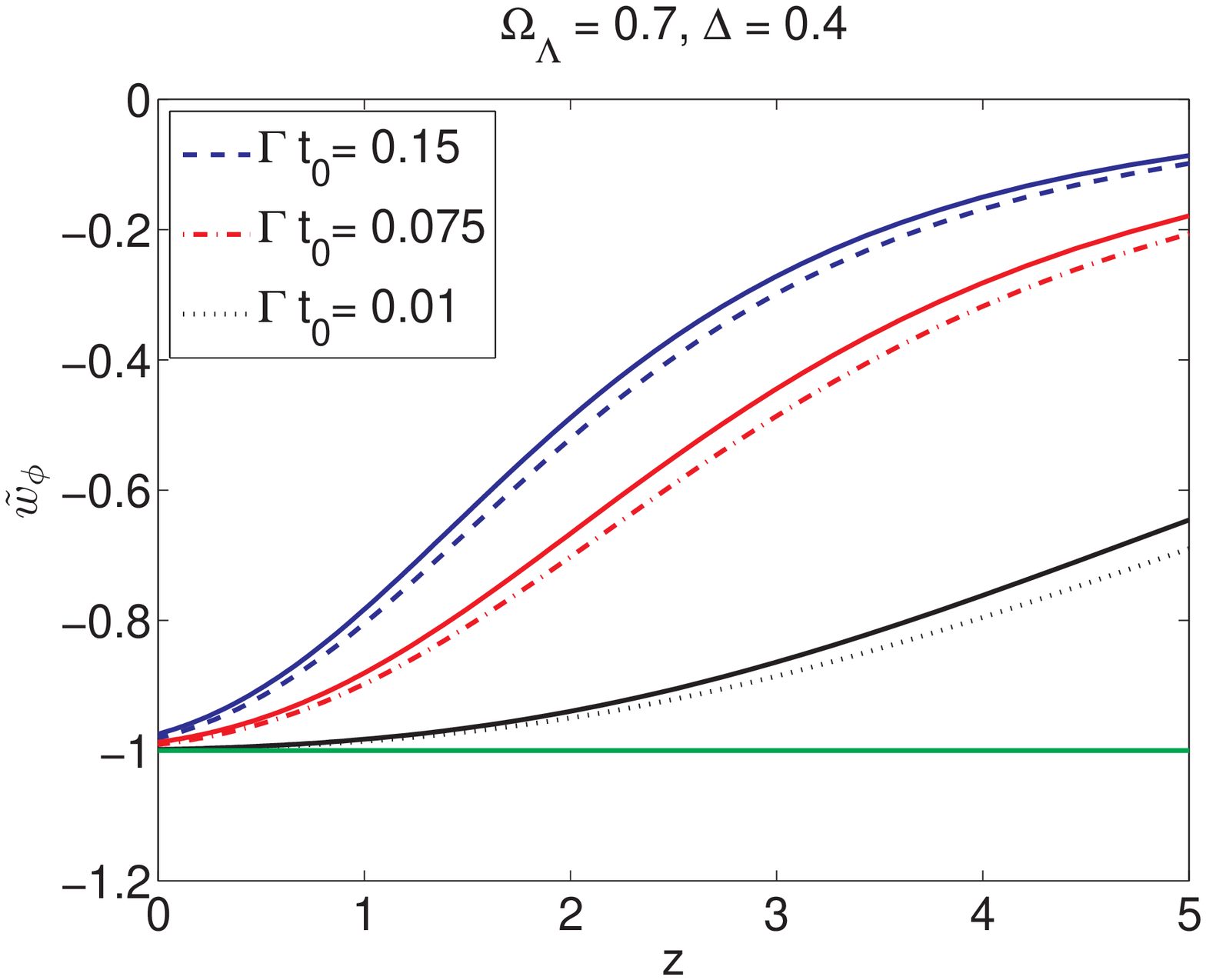,width=7 cm}\\
\epsfig{file=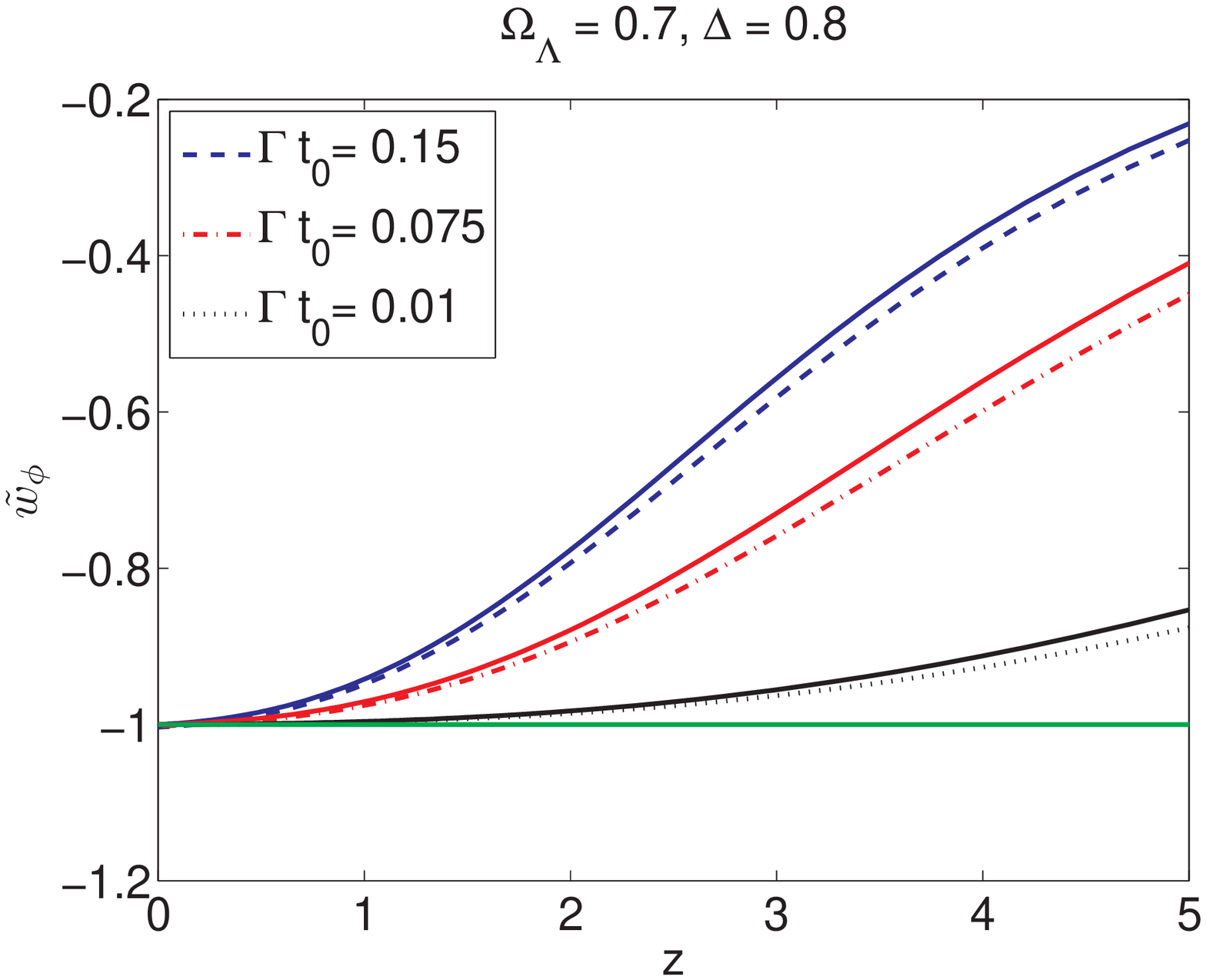,width=7 cm}&\epsfig{file=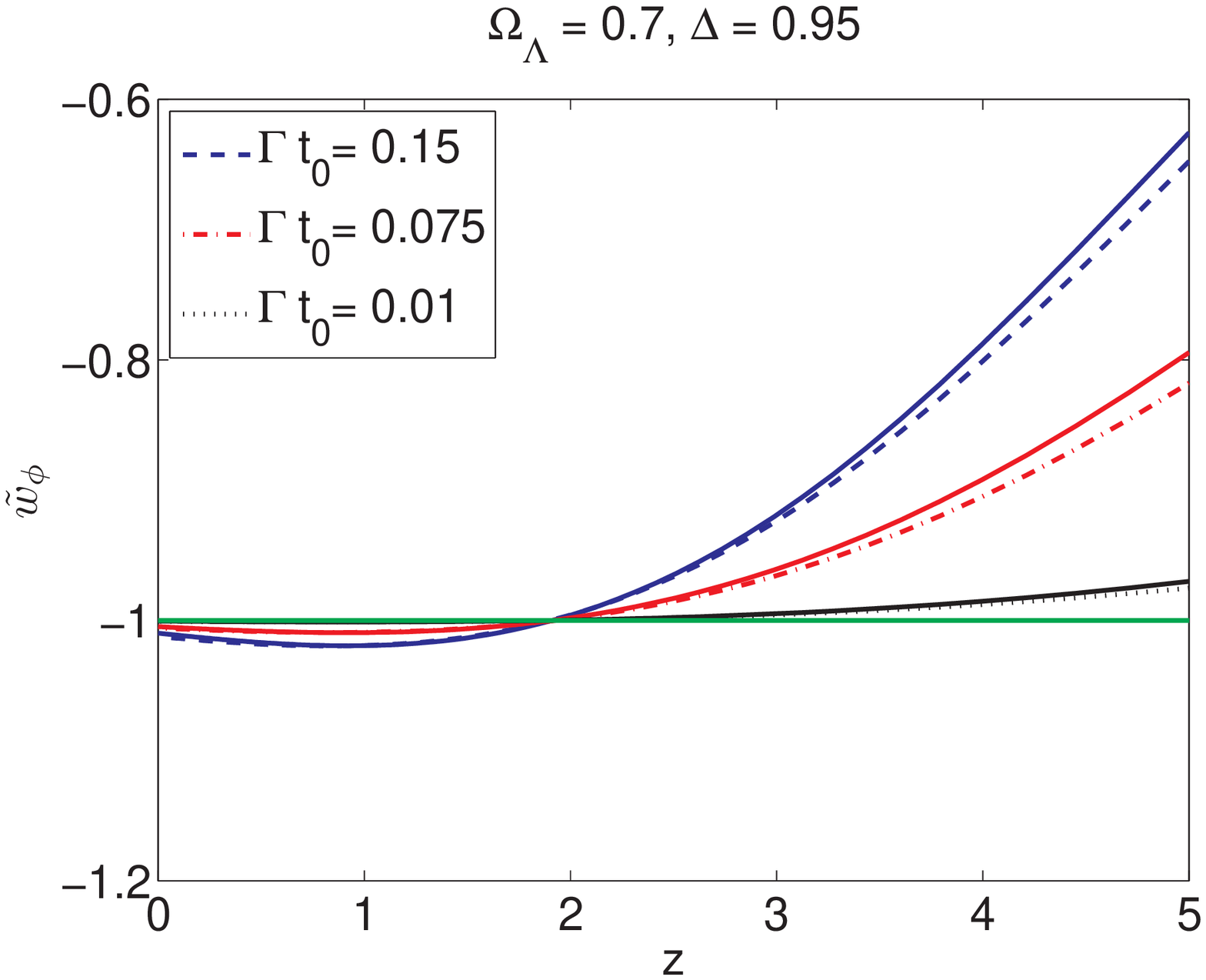,width=7
cm}\\
\epsfig{file=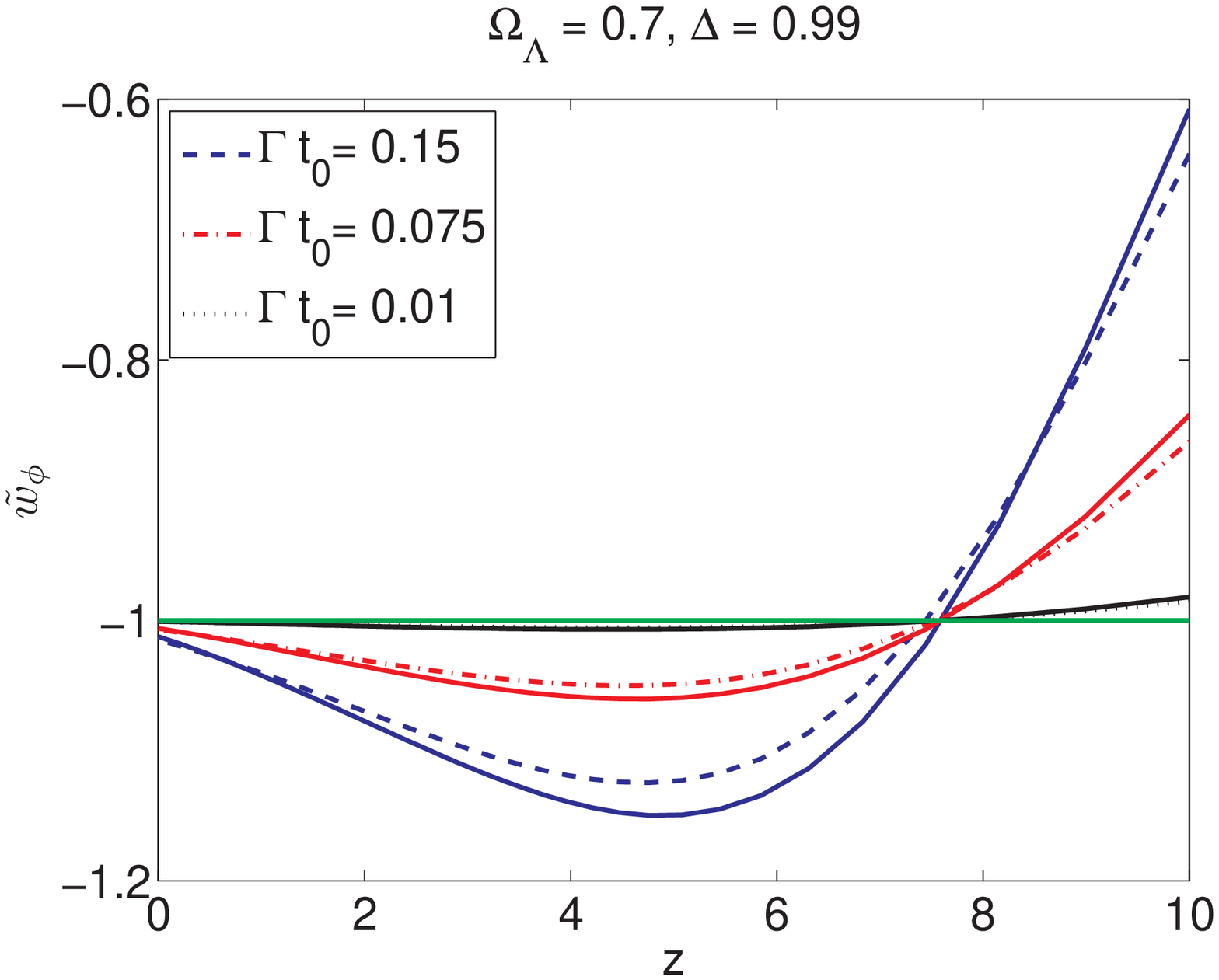,width=7cm}&\epsfig{file=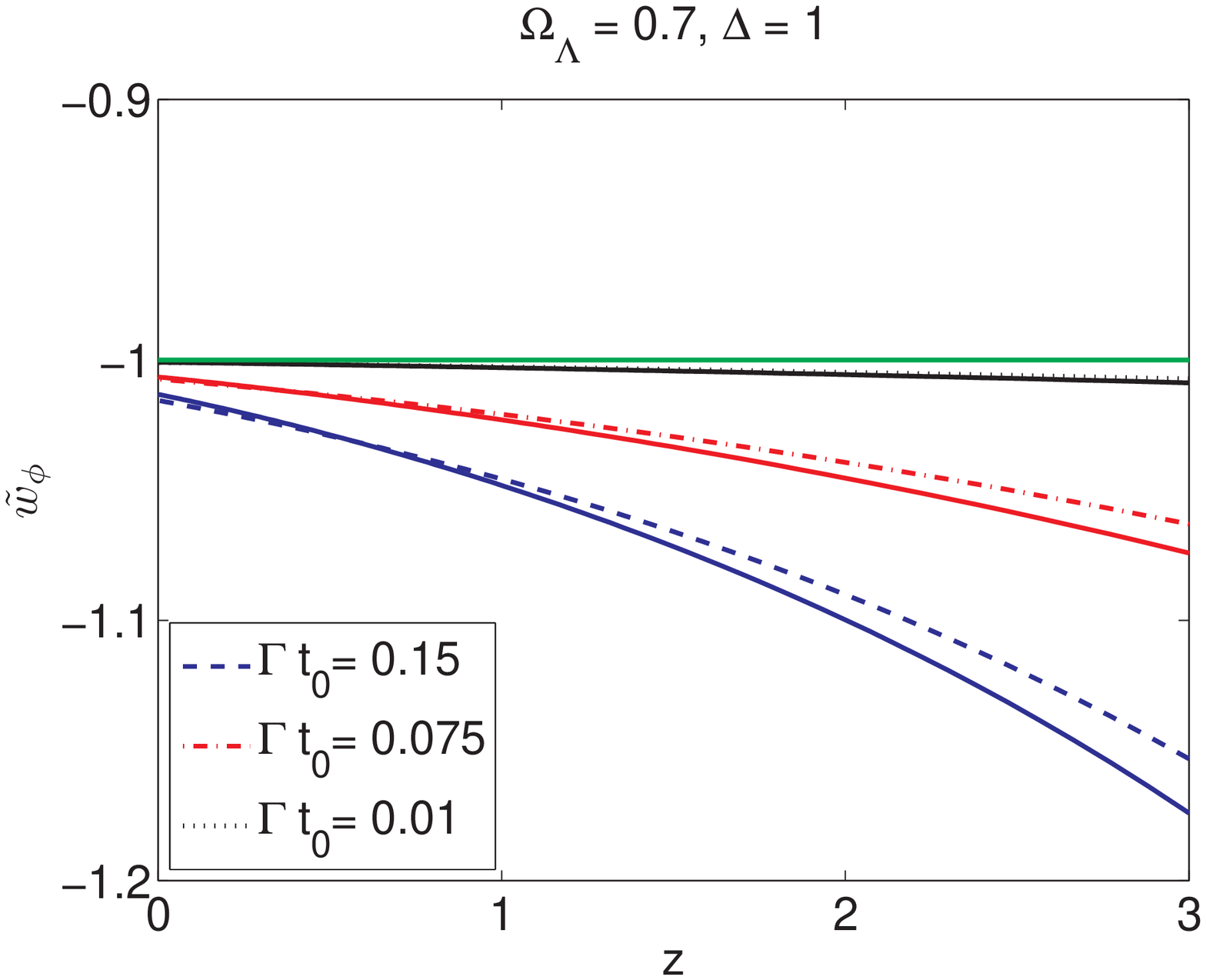,width=7cm}
\end{tabular}
\caption{\label{wz_varying_Gammas}
The evolution of the spurious equation of state parameter, $\widetilde w_\phi$,
in a $\Lambda$CDM model ($\Omega_\Lambda = 0.7$) with decaying dark matter, in which the dark matter
is erroneously taken to be stable and the corresponding time variation is
absorbed into a time-varying equation of state for the dark energy.
Here $\Gamma$ is the decay rate of the dark matter, $t_0$ is the age
of the universe, and $\Delta$ is determined by the assumed
value
of the dark matter density, erroneously taken to scale as $a^{-3}$.
Broken lines denote the exact evolution and solid
lines denote the analytic approximation given by Eqs. (\ref{wfinal})-(\ref{fz}).}
\end{figure*}
\end{center}

\begin{center}
\begin{figure*}[!]
\begin{tabular}{c@{\qquad}c}
\epsfig{file=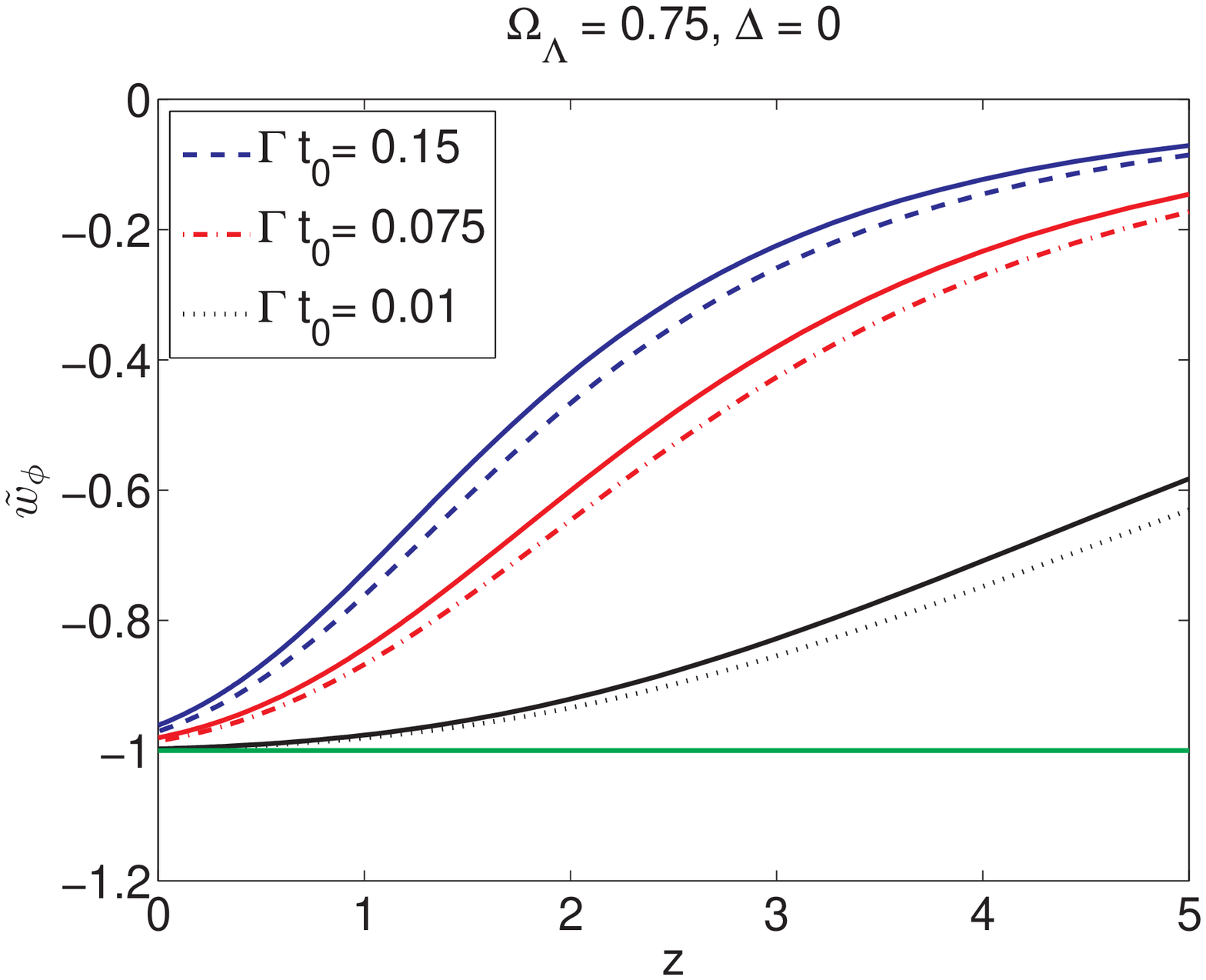,width=7 cm}&\epsfig{file=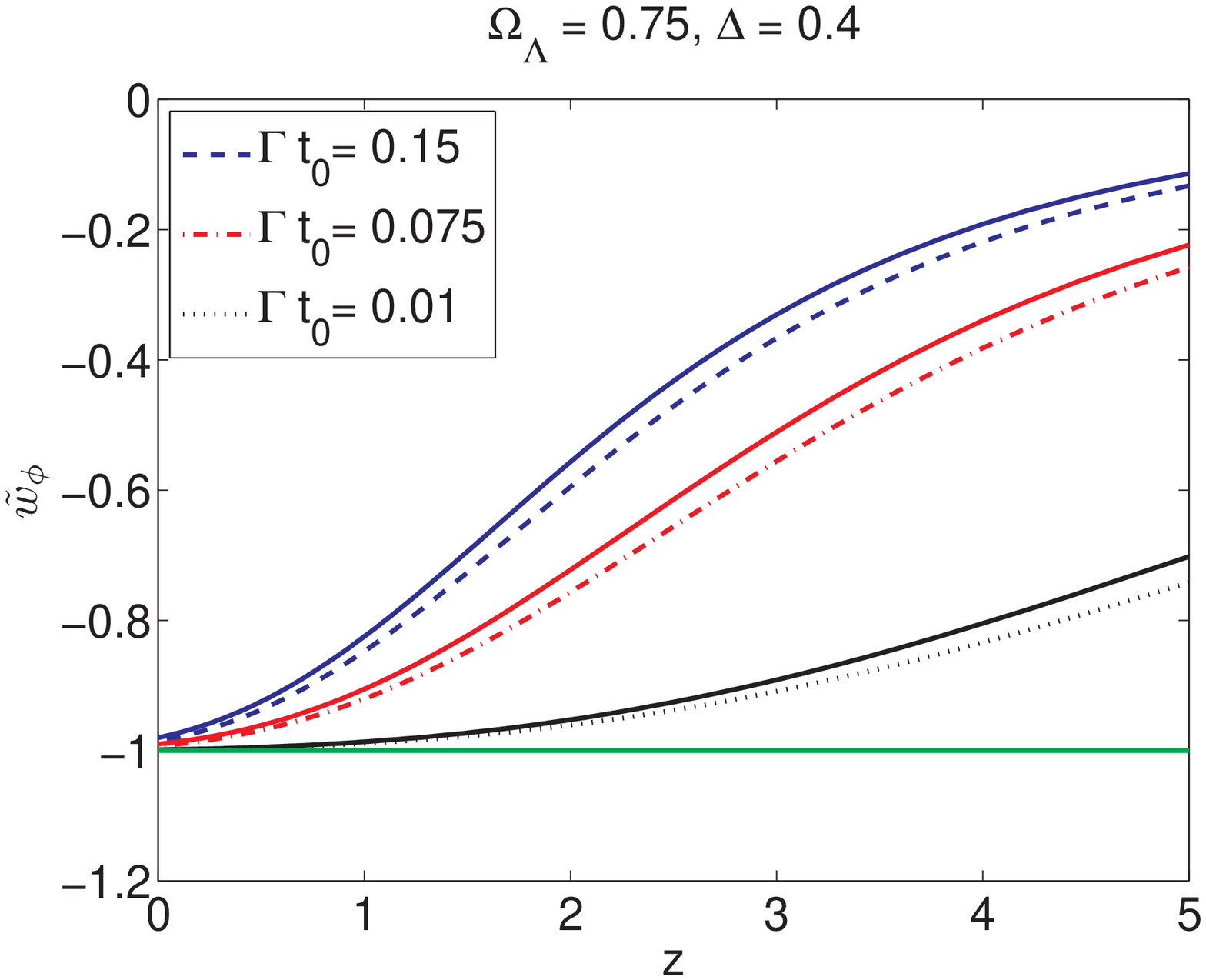,width=7 cm}\\
\epsfig{file=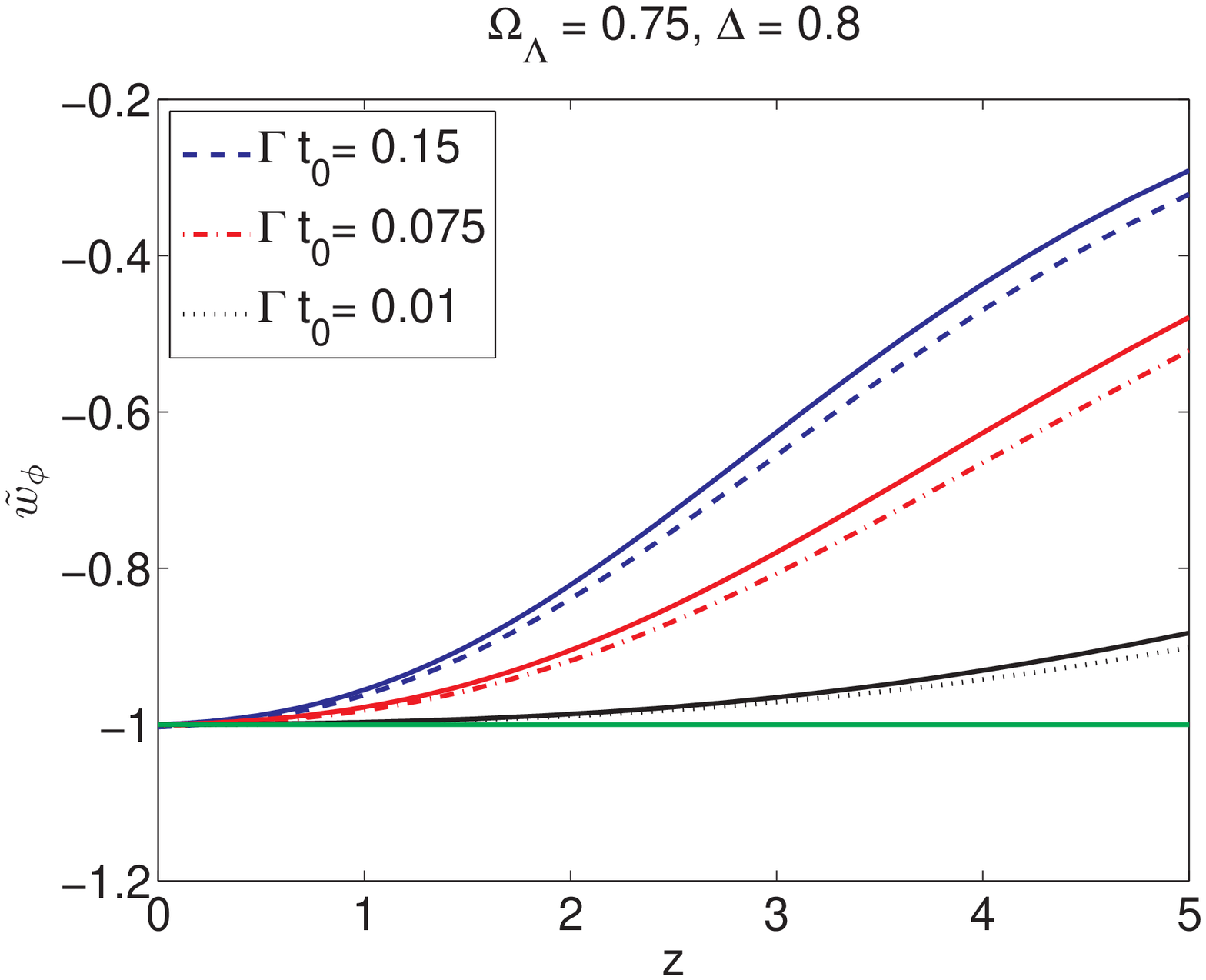,width=7 cm}&\epsfig{file=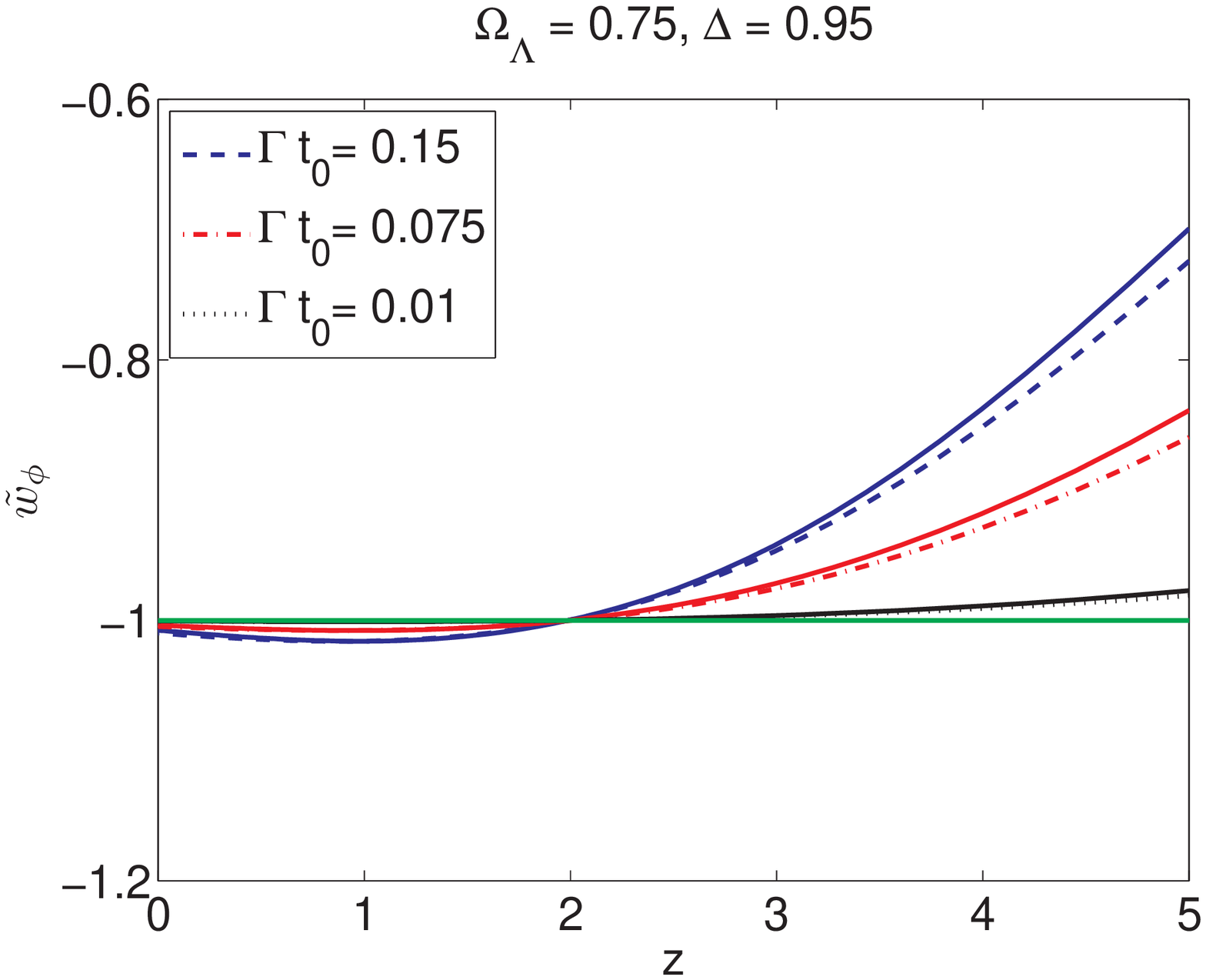,width=7
cm}\\
\epsfig{file=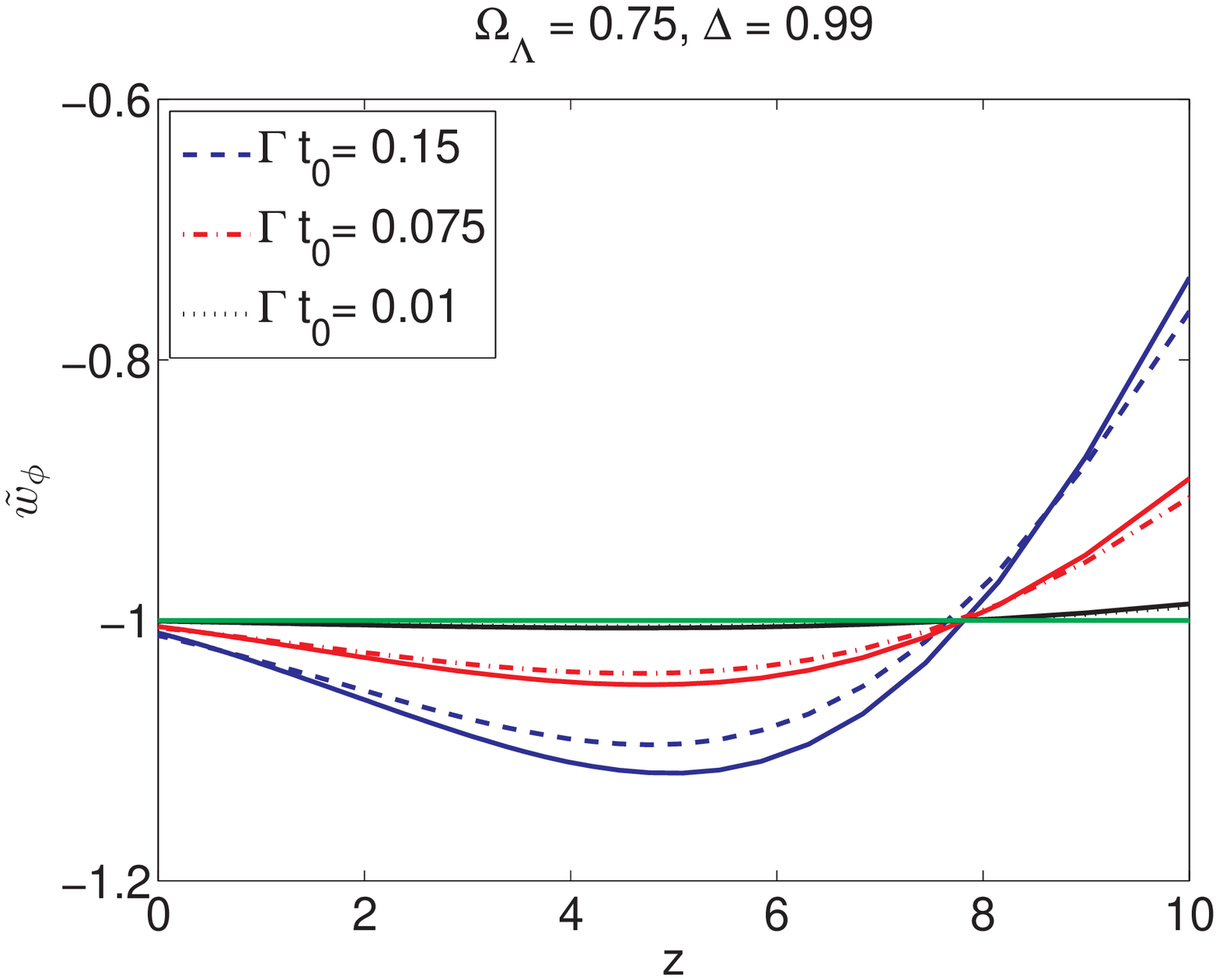,width=7cm}&\epsfig{file=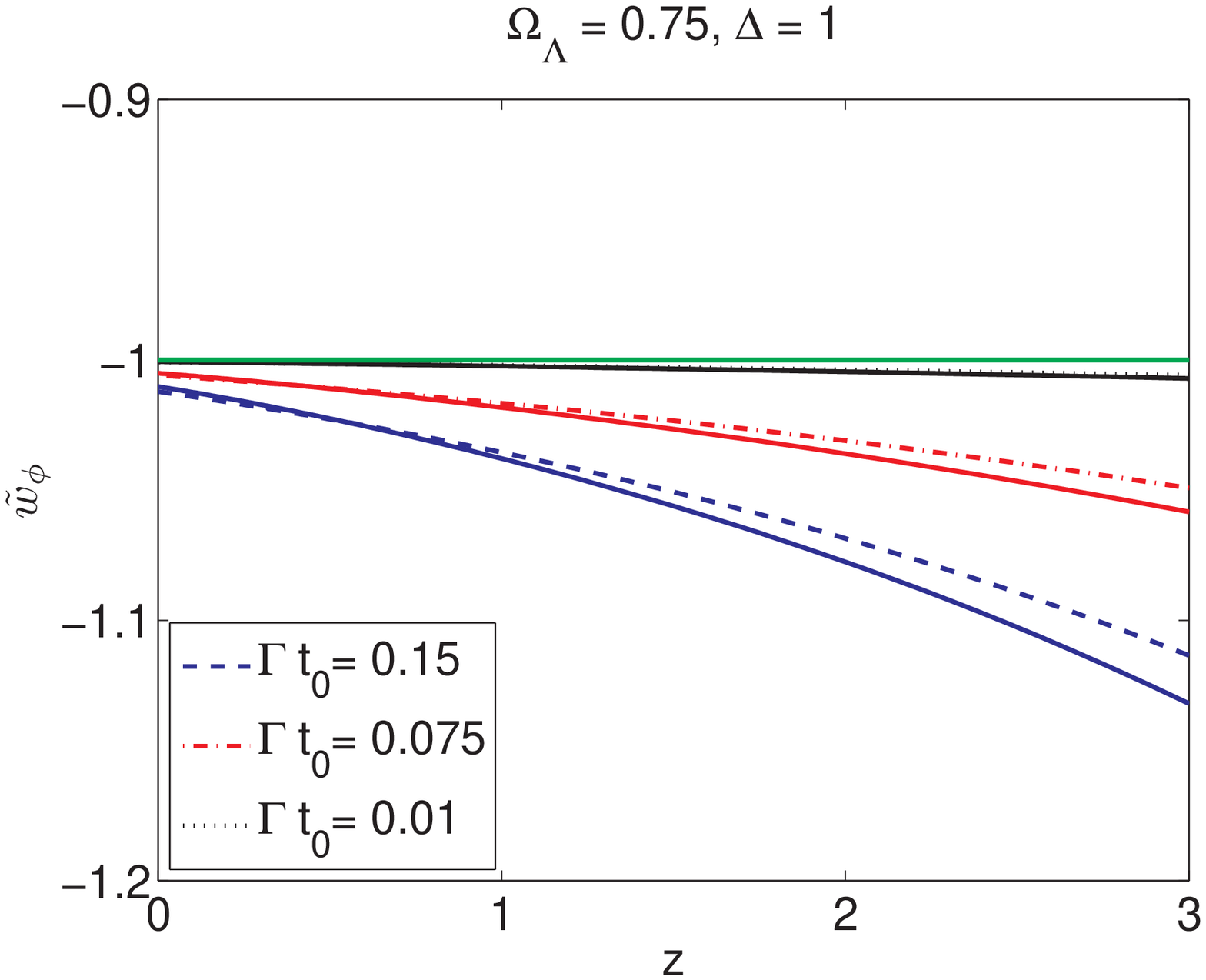,width=7cm}
\end{tabular}
\caption{As Fig. 1, with $\Omega_\Lambda = 0.75$.}
\end{figure*}
\end{center}

\section{Discussion}

Eqs. (\ref{wfinal}) and (\ref{fz}) yield several important insights
into the behavior of $\widetilde w_\phi$.  First, we note that all of our
models converge to $\widetilde w_\phi \rightarrow -1$ at late times.  This
is obvious from the fact that $\rho_\Lambda$ is always the dominant
component at late times.  However, the way in which $\widetilde w_\phi$ evolves
with redshift is extremely sensitive to the assumed value of
$\widetilde \rho_{DM}$ through its dependence
on $\Delta$.  For small values of $\Delta$ (i.e., taking $\widetilde
\rho_{DM}$ nearly equal to the present-day value of $\rho_{DM}$) the spurious
time-varying dark energy evolves as a quintessence component with $\widetilde
w_\phi > -1$, with $\widetilde w_\phi$ decreasing with time.  In
the terminology of Ref. \cite{CL}, the model mimics a ``freezing" quintessence
field.

For larger values of $\Delta$, the model crosses the
phantom divide at some redshift $z_c$, with $\widetilde w_\phi > -1$
at $z > z_c$ and $\widetilde w_\phi < -1$ at $z < z_c$.
An approximate value for $z_c$ can be derived from Eq. (\ref{wfinal}):
\be
\label{zstar}
1+z_c= \left(\frac{\Omega_\Lambda}{(1-\Omega_\Lambda)
\sinh^2[5(1-\Delta)\tanh^{-1}(\sqrt{\Omega_\Lambda})]}\right)^{1/3}.
\ee
Note that $z_c$ is independent of $\Gamma$ (which is apparent in Figs. 1-2)
and depends only on $\Delta$.  
The expression for the time $t_c$ at which $\widetilde w_\phi$ crosses $-1$ can
be derived from Eq. (\ref{w_ana}) and takes the particularly simple form
\be
\label{tc}
t_c/t_0 = 5(1-\Delta).
\ee
From Eq. (\ref{tc}), it is obvious that these
models cross the phantom divide whenever $0.8 < \Delta < 1$.
(Strictly speaking, models with $\Delta < 0.8$
also cross the phantom divide, but this crossing takes place in the future,
while $\Delta = 0.8$ gives $\widetilde w_\phi$ exactly equal to $-1$ at the present).
Finally, for $\Delta = 1$ (i.e., setting $\widetilde \rho_{DM}$ equal
to $\rho_{DM}$ at early times) the model behaves like a phantom,
with $\widetilde w_\phi < -1$ at all times.


The $\Lambda$CDM model with decaying dark matter can be distinguished
from the corresponding quintessence model with the same $H(z)$ using
the growth of density perturbations.  (For a general discussion of linear
perturbation growth as a probe of dark energy, see, e.g., Refs.
\cite{caldwelletal,ma,KLSW,LinderJenkins,Linder,Ishak1,Dutta_pert1,Dutta_pert2,
Dutta_pert3,Dutta_pert4,Ishak2,Dutta_pert5}).
The equation for the evolution of the density perturbation, $\delta$,
in the linear regime ($\delta \ll 1)$, well inside the horizon, is
\begin{equation}
\label{delta}
\ddot \delta + 2H \dot \delta - 4 \pi G (\rho_{DM} + \rho_B) \delta = 0,
\end{equation}
where the dot denotes the derivative with respect to time, and
we assume
the baryons are decoupled from the photon background and can cluster freely.
By construction, the value of $H(z)$ is the same
in the $\Lambda$CDM model with decaying dark matter and in
the corresponding spurious quintessence model with stable dark matter.
However, these two models differ in the value of $\rho_{DM}$,
which drives the growth of density perturbations in equation (\ref{delta}).  In the
$\Lambda$CDM model with decaying dark matter, $\rho_{DM}$ is given
by Eq. (\ref{rhoM_sol}),
while the corresponding spurious quintessence model has a (spurious)
matter density $\widetilde \rho_{DM}$ given by Eq. (\ref{rhom_sol}). 
Then using equation (\ref{Deltadef}), we see that
\begin{equation}
\frac{\widetilde \rho_{DM}}{\rho_{DM}} = \exp[\Gamma t +
(\Delta-1)\Gamma t_0].
\end{equation}
The growth rate in the spurious quintessence model will be larger (smaller)
than the corresponding growth rate in the $\Lambda$CDM model as
$\Gamma t + (\Delta-1)\Gamma t_0$ is greater than (less than) zero.  In the
limiting case $\Delta = 1$ the growth rate will be
larger for the spurious quintessence model than for the decaying dark
matter model over the entire evolution history, while the reverse is true
for $\Delta = 0$.  Independent of $\Delta$, the ratio of the perturbation
growth rate at late
times to the growth rate at early times is smaller for
the decaying dark matter model
than for the corresponding spurious quintessence model with the same $H(z)$.  This result
follows from the fact that the dark matter density decreases more rapidly with
scale factor in the former model.  

An obvious question is whether or not these results can ever be relevant.
If dark matter really were unstable, with a lifetime much longer than
the age of the universe, then it is certainly plausible that a
cosmological
signature of unstable dark matter
would be detected first in some combination of CMB and large-scale
structure observations.  In this case, the information on decaying dark
matter would simply be incorporated into calculations for the equation of
state of the dark energy, rendering our calculations moot.  On the other
hand, it is also plausible that such effects would be detected first
in precision measurements of the dark energy equation of state, in which
case our expressions for $\widetilde w_\phi$ provide a useful guide to
the sort of spurious signal produced by decaying dark matter.

These results can be generalized, in an obvious way, to quintessence models (or other dark
energy models with a time-varying equation of state) with unstable dark matter,
to derive the corresponding change in the measured equation of state parameter.  Such models,
however, must be considered somewhat baroque.

\section{Acknowledgments}
R.J.S. was supported in part by the Department of Energy
(DE-FG05-85ER40226).  We thank H. Ziaeepour for helpful discussions.


\begin{thebibliography}{99}

  
\bibitem{hicken}
M.~Hicken {\it et al.},
  arXiv:0901.4804 [astro-ph.CO].
  

\bibitem{Komatsu:2010fb}
  E.~Komatsu {\it et al.}  [WMAP Collaboration],
  Astrophys.\ J.\ Suppl.\  {\bf 180}, 330 (2009)
  [arXiv:0803.0547 [astro-ph]].


\bibitem{percival}
  W.~J.~Percival {\it et al.},
  Mon.\ Not.\ Roy.\ Astron.\ Soc.\  {\bf 401}, 2148 (2010)
  [arXiv:0907.1660 [astro-ph.CO]].


\bibitem{Samushia2007}
  L.~Samushia, G.~Chen and B.~Ratra,
  arXiv:0706.1963 [astro-ph].


\bibitem{Ettori}
  S.~Ettori {\it et al.},
  arXiv:0904.2740 [astro-ph.CO].






\bibitem{Wang}
  Y.~Wang,
  Phys.\ Rev.\  D {\bf 78}, 123532 (2008).
  
\bibitem{Samushia2009}
  L.~Samushia and B.~Ratra,
  arXiv:0905.3836 [astro-ph.CO].


\bibitem{Copeland}
E.J. Copeland, M. Sami, and S. Tsujikawa, Int. J. Mod. Phys. D
{\bf 15}, 1753 (2006).

\bibitem{Wood-Vasey}
W.M. Wood-Vasey, et al., \apj {\bf 666}, 694 (2007).

\bibitem{Davis}
T.M. Davis, et al., \apj {\bf 666}, 716 (2007).

\bibitem{ratra}
B. Ratra and P.J.E. Peebles,
\prd {\bf 37}, 3406 (1988).

\bibitem{wett88}
C. Wetterich, Nucl. Phys. B {\bf 302}, 668 (1988) 

\bibitem{turner}
M.S. Turner and M. White,
\prd {\bf 56}, R4439 (1997).

\bibitem{caldwelletal}
R.R. Caldwell, R. Dav{\'e}, and P.J. Steinhardt,
\prl {\bf 80}, 1582 (1998).

\bibitem{liddle}
A.R. Liddle and R.J. Scherrer,
\prd {\bf 59}, 023509 (1999).

\bibitem{Stein1}
P.J. Steinhardt, L. Wang, and I. Zlatev,
\prd {\bf 59}, 123504 (1999).

\bibitem{DS}
  S.~Dutta and R.~J.~Scherrer,
  Phys.\ Rev.\  D {\bf 78}, 123525 (2008);
  S.~Dutta, E.~N.~Saridakis and R.~J.~Scherrer,
  Phys.\ Rev.\  D {\bf 79}, 103005 (2009);
    S.~Dutta and R.~J.~Scherrer,
  Phys.\ Lett.\  B {\bf 676}, 12 (2009).
  
  
  \bibitem{Dutta1}
  S.~Dutta and R.~J.~Scherrer,
  Phys.\ Rev.\  D {\bf 78}, 083512 (2008);
  \bibitem{Dutta2}
  S.~Dutta, S.~D.~H.~Hsu, D.~Reeb and R.~J.~Scherrer,
  Phys.\ Rev.\  D {\bf 79}, 103504 (2009).
  

\bibitem{Hu}
W. Hu and D.J. Eisenstein, \prd {\bf 59},
083509 (1999).

\bibitem{Wasserman}
I. Wasserman, \prd {\bf 66},
123511 (2002).

\bibitem{Rubano}
C. Rubano and P. Scudellaro, Gen. Rel. Grav.
{\bf 34}, 1931 (2002).

\bibitem{Kunz}
M. Kunz, \prd {\bf 80}, 123001 (2009).

\bibitem{Z}
H. Ziaeepour, astro-ph/0002400.

\bibitem{Das}
S. Das, P.S. Corasaniti, and J. Khoury,
\prd {\bf 73}, 083509 (2006).

\bibitem{Kaplinghat}
M. Kaplinghat, R.E. Lopez, S. Dodelson, and R.J. Scherrer,
\prd {\bf 60}, 123508 (1999).

\bibitem{Ichiki}
K. Ichiki, M. Oguri, and K. Takashai, \prl {\bf 93},
071302 (2004).

\bibitem{Gong}
Y. Gong and X. Chen, arXiv:0802.2296.

\bibitem{Amigo}
S.D. Amigo, W. M.-Y. Cheung, Z. Huang, and S.-W. Ng,
JCAP {\bf 6}, 005 (2009).

\bibitem{ST}
R.J. Scherrer and M.S. Turner, \prd {\bf 31}, 681 (1985).


\bibitem{CL}
R.R. Caldwell and E.V. Linder, \prl {\bf 95}, 141301 (2005).

\bibitem {ma} C.-P. Ma, R.R. Caldwell, P. Bode, and L. Wang,
\apj {\bf 521}, L1 (1999).

\bibitem{KLSW} J. Kujat, A.M. Linn, R.J. Scherrer, and D.H. Weinberg,
\apj {\bf 572}, 1 (2002).

\bibitem{LinderJenkins} E.V. Linder and A. Jenkins, 
Mon. Not. R. Astron. Soc. {\bf 346}, 573
(2003).

\bibitem{Linder} E.V. Linder, \prd {\bf 72}, 043529 (2005).

\bibitem{Ishak1} M. Ishak, A. Upadhye, and D.N. Spergel,
\prd {\bf 74}, 043513 (2006).



\bibitem{Dutta_pert1}
  S.~Dutta and I.~Maor,
  Phys.\ Rev.\  D {\bf 75}, 063507 (2007)



\bibitem{Dutta_pert2}
  J.~B.~Dent, S.~Dutta and T.~J.~Weiler,
  Phys.\ Rev.\  D {\bf 79}, 023502 (2009)

\bibitem{Dutta_pert3}
  J.~B.~Dent and S.~Dutta,
  Phys.\ Rev.\  D {\bf 79}, 063516 (2009)


\bibitem{Dutta_pert4}
  J.~B.~Dent, S.~Dutta and L.~Perivolaropoulos,
  Phys.\ Rev.\  D {\bf 80}, 023514 (2009)

\bibitem{Ishak2} J. Dossett, et al., JCAP {\bf 1004},
022 (2010).


\bibitem{Dutta_pert5}
  J.~C.~Bueno Sanchez, J.~B.~Dent, S.~Dutta and L.~Perivolaropoulos,
  arXiv:1004.4905 [astro-ph.CO].

\end{thebibliography}
\end{document}